# A Near-Infrared Variant of the Barnes-Evans Method For Finding Cepheid Distances Calibrated with High-Precision Angular Diameters

DOUGLAS L. WELCH[1]

Department of Physics and Astronomy, McMaster University,
Hamilton, Ontario L8S 4M1 CANADA

astro-ph/9407088  18 Aug 94

Address for proofs: Dr. Douglas L. Welch
Department of Physics and Astronomy
McMaster University
1280 Main Street West
Hamilton, Ontario
L8S 4M1
CANADA

---

# ABSTRACT

The advantages of a near-infrared variant of the Barnes-Evans method for estimating distances to Cepheid variables are described and quantified. A surface brightness-color relation for $K$ photometry and the $(V-K)_0$ color index is established using modern, high-precision angular diameters from optical interferometers. Applied to data for the galactic (cluster) Cepheid U Sgr, this method yields a distance of $0.660 \pm 0.024$ kpc and a true distance modulus of $9.10 \pm 0.07$ mag. This estimate compares with the true distance modulus of $9.37 \pm 0.22$ mag estimated by Gieren, Barnes, and Moffett (1993) using the classical Barnes-Evans technique. The possibility of estimating distances of LMC and SMC Cepheids directly – without intermediate steps – is discussed. The feasibility of determining the distance of M31 or M33 using this technique is examined and is probably within the reach of 8m-class telescopes.

*Key Words*: Stars – Intrinsic Variables; Stars – Variable Stars; Stellar Systems – Open Clusters; Stellar Systems – Globular Clusters; Stellar Systems – Magellanic Clouds

*Suggested Running head*: High-Precision Cepheid Distance Scale

## 1. Introduction and Motivation

The calibration of primary distance indicators such as classical Cepheid variables remains a major source of uncertainty in the extragalactic distance scale (Jacoby *et al.* 1992). It is clearly desirable to employ independent calibration techniques to ensure that unexpected systematic errors in derived distances have the possibility of being detected. To date, a great deal of trust has been placed in the calibration of Cepheids in Galactic open clusters using main-sequence fitting techniques.

The recent availability of high signal-to-noise $JHK$ photometry and precise radial velocities for Galactic and Magellanic Cloud Cepheids now allows for more accurate radius determination than has hitherto been possible. When combined with a precise surface brightness, angular diameter calibration, geometric distances to Cepheids become possible which are completely independent of the main-sequence fitting calibration. As shown in this paper, it is now possible to derive distances accurate to a few percent out to the Magellanic Clouds based on a variant of the Barnes-Evans technique which has traditionally employed $V$ photometry and the $(V-R)_0$ color index.

## 2. The Barnes-Evans Method

Like any geometric method, the Barnes-Evans technique combines actual physical displacements (obtained by integrating the radial velocity curve) with angular diameters (found with a surface brightness-color relation) to infer the distance and mean radius which makes those two sets of numbers most consistent. The technique was first employed by Barnes *et al.* (1977). In that paper, the surface brightness was defined for the $V$ bandpass and the color employed was $(V-R)_0$. This choice of color index followed examination of the surface brightness behavior of other indices, including $(U-B)_0$, $(B-V)_0$, and $(R-I)_0$ in Barnes and Evans (1976), Barnes, Evans, and Parsons (1976) and Barnes, Evans, and Moffett (1978). The index $(V-R)_0$ was judged to have the best combination of insensitivity to surface gravity and sensitivity to temperature of the four studied. Recently, Gieren, Barnes, and Moffett (1993) have estimated the distances of 100 galactic Cepheids using the Barnes-Evans technique. They conclude that there is no systematic difference between cluster main sequence fitting distances and Barnes-Evans distances.

The surface brightness relation for V and $(V-R)_0$ can be determined from first principles as described by Barnes and Evans (1976) and later workers. That is, the zeropoint of the relation can potentially be derived knowing the bolometric correction, effective temperature, apparent magnitude and apparent diameter of the Sun. The slope of the surface brightness-color relation can be predicted from the synthetic photometry of model atmosphere fluxes. However, a second approach is possible. The surface brightness-color relation can be defined entirely in terms of observations of stars of known angular diameter. In the past, this approach was not obviously the most desirable because the angular diameters were only well-determined for upper main sequence stars from intensity-interferometer measurements (see Wesselink 1969 and Hanbury-Brown *et al.* 1967). The angular diameters for cool stars were principally from lunar occultations. Unfortunately, there are few stars that can be used to calibrate the relation in the color interval containing Cepheid variables. This situation has recently changed with the first published results of the Mark III Optical Interferometer on Mount Wilson by Mozurkewich *et al.* (1991) which presents 2-3% angular diameters for 12 intermediate color stars allows the first meaningful attempt to calibrate the surface brightness-color relation directly.

### 3. Advantages of the Near-Infrared

To extract the variation in stellar radius from the lightcurve of a Cepheid, it is highly desirable to select a color index to infer the temperature variation which is very sensitive. Although 'blue' color indices such as $(U-B)_0$, $(B-V)_0$, or $(U-V)_0$ have high temperature sensitivity, they are also strongly affected by line-blanketing and lines sensitive to surface gravity, so they should not be used. The other way to ensure temperature sensitivity is to employ an index with a long wavelength baseline. Perhaps the ideal index is $(V-K)_0$ – both bandpasses are only lightly line-blanketed and precisions of better than 1% are routinely possible in the photometry in each bandpass (a fact not true for redder bandpasses such as $L$' or $M$). Furthermore, the $K$ bandpass does not sit on a major opacity feature such as the $H^-$ opacity minimum at 1.6 $\mu$m.

The choice of the 'flux' bandpass is equally important. $K$ is almost ideal. Since it is located in the Rayleigh-Jeans tail of the flux distribution of Cepheids and hence reflects far less of the temperature variation than a bluer bandpass. At 2.2 $\mu$m the flux variation due to radius change is typically comparable to that due to effective temperature variation.

In short, we pick a bandpass which is insensitive to temperature variation in the first place and remove the effect of temperature change with a very sensitive color index! We will presently illustrate the relative power of the $K$,$(V-K)_0$ data compared to the combination $V$,$(V-R)_0$.

## 4. The Calibration of $S_K((V-K)_0)$

Before proceeding, it is necessary to define a surface brightness-color relation. To do so, we require $V$ and $K$ photometry and angular diameters. The latter will be taken entirely from Mozurkewich *et al.* (1991) for the sake of homogeneity, even though there are occasional additional sources such as Di Benedetto and Rabbia (1987). The angular diameters of Mozurkewich *et al.* need to be corrected to the expected uniform disk diameter at $K$. We have used the limb-darkening coefficients of Parsons (1971) and Manduca (1979) for this purpose. We adopt the value of $\theta_{LD}$ from Table 3 of Mozurkewich *et al.* (which represents the limb-darkened angular diameter) and correct it to the 2.2 $\mu$m uniform disk diameter $\theta_{UD}^K$ by:

$$\theta_{UD}^{K} = \sqrt{1 - \frac{U}{3} - \frac{V}{6}}\ \theta_{LD},$$

where U and V are tabulated in Parsons (1971). Their direct counterparts in Manduca (1979) are $A'$ and $B'$. The range in the correction factor is small. For the models in Parsons at $\log g = 1.2$, the factor varies from 0.963 at 5400 K to 0.969 at 6600 K. Similarly, for solar abundance star at $\log g = 1.5$ in Manduca, the factor varies between 0.954 at 4500 K to 0.950 at 3750 K. As a simple interpolation for this factor, we adopt a correction factor, $k$, given by:

$$k = 0.975 - 0.005\ (V - K)_0.$$

Table 1 contains pertinent observational information and references. The columns are, from left to right, star designation, spectral type, limb-darkened angular diameter from Table 3 of Mozurkewich *et al.* (1991) in milliarcseconds (mas), derived uniform-disk angular diameter in mas expected at $K$, $K$ magnitude, $(V - K)_0$ color index in magnitudes, and surface brightness at $K$ which is equivalent to the magnitude for a star of the same color with an angular diameter of 1 mas. The uniform disk diameters at $K$ contain the limb-darkening correction factor, $k$, described above. $K$ magnitudes are taken from the "Provisional KPNO IR Standards" list of R. Joyce (private communication) for the five stars $\alpha$ Ari, $\alpha$ CMi, $\beta$ Gem, $\epsilon$ Cyg, and $\beta$ Peg. The remaining $K$ magnitudes listed (with the exception of $\alpha$ Cet) are a straight average of the values listed in Gezari *et al.* (1993), where the means were determined from 2 to 26 measurements, depending on the star. While this expedient is not ideal by any means, it is probably superior to the adoption of a few individual studies which have unknown transformation properties to a standard system. (A comparison of the values from the Joyce list and the Gezari averages resulted in a mean zeropoint difference of 0.003 mag and a maximum difference of 0.025 mag for the five stars in common.) $\alpha$ Cet was observed by Dana Backman and Sergio Fajardo-Acosta on the CTIO 1.5m telescope on 1993 November 28 using the D3 InSb infrared photometer with a 9.7 arcsec aperture. V photometry is from Johnson *et al.* (1966). *The error limiting the precision of the calibration is probably the quality of the V and K photometry of bright stars.*

Figure 1 is a plot of $S_K$ versus $(V - K)_0$, where the data from Table 1 are used, and

$$S_K = K + 5 \log_{10} \theta_{UD}^K.$$

Also shown is an unweighted least-square line fit through the points, valid only within the range of colors of the calibrating stars $(1 < (V - K)_0 < 6)$. This relationship has a $\sigma = 0.04$ mag, consistent with all of the scatter being due to photometric errors and expected uncertainties in the angular diameters. It has been assumed that all stars are close enough that $E_{V-K}$ in negligible. The equation for this line is:

$$S_K = (2.76 \pm 0.04) + (0.252 \pm 0.012)(V - K)_0.$$

Naively, one might expect that the use of $(V - K)$ would result in a tremendous sensitivity to uncertainties in the total absorption. In fact, the infrared Barnes-Evans technique has reddening properties similar to its optical counterpart – for the same reason: that temperature and reddening affect the color and brightness is almost the same way. So if the assumed absorption is too low, the color adopted will be too red, and this will infer a fainter surface brightness. The magnitude inserted into $S_K$ will be fainter than the true unreddened magnitude, but the surface brightness derived from the incorrect color was also too faint, so the error in the angular diameter derived is partially reduced. The actual sensitivity will be derived in the next section where we give an example reduction based on real data.

## 5. A Galactic Cluster Cepheid: U Sgr

The power of the near-infrared Barnes-Evans technique is clearly illustrated using real data. Since we know that the angular diameters are good to 2-3%, we require high-precision photometry to ensure that the derived distance and radius are not degraded further. A Galactic Cepheid with extensive, high-precision photometry and radial velocities is U Sgr. This star is also a member of the open cluster M25 (=IC 4725) and hence has independent estimates available. $V$ photometry (36 observations) is taken from Moffett and Barnes (1984), where the expected uncertainty is about 0.005 mag. $K$ photometry (30 points) is found in Laney and Stobie (1992) where the uncertainty is thought to be 0.010 mag. (The $K$ zeropoint of the both the Joyce standard list and the Carter (1990) system used by Laney and Stobie are equivalent to Vega having $K$ = 0.00 mag.) Radial velocities are from Mermilliod, Mayor, and Burki (1987) which contains 42 observations with uncertainties of 0.5 km s$^{-1}$.

The radial velocity curve must be integrated to obtain the displacement from mean radius. To do so, we require a projection correction factor, $p$, to account for the fact that observed radial velocity of the photosphere is always less than the actual expansion (or contraction) velocity due to the contributions of surface elements with components of their motion which are not along the line-of-sight. Following Hindsley and Bell (1986), we adopt $p = 1.36$ as being suitable for the CORAVEL velocities used in this paper.

A complication of using both an optical and an infrared flux to define the color index is that the measurements are not usually obtained simultaneously – that is, few observatories are equipped with combination optical/infrared photometers. To define a color index at the time of each $K$ measurement, it is necessary to fit the $V$ lightcurve and predict a $V$ magnitude for the phase of each $K$ datum. A fifth-order Fourier series was found to provide a fit with $\sigma = 0.018$ mag for the $V$ measurements available. For a Fourier series of the form:

$$V = a_0 + \sum_{n=1}^{5} a_n \cos(2n\pi\phi) + b_n \sin(2n\pi\phi),$$

where the phase, $\phi$, is that adopted by Mermilliod, Mayor, and Burki (1987):

$$\phi = \frac{\text{Frac}(\text{RJD} - 45871.660)}{6.7453}.$$

The values of the Fourier coefficients found from the fit are:

$$
\begin{aligned}
a_0^V &= 6.7198, \\
a_1^V &= 0.2485, & b_1^V &= 0.1752 \\
a_2^V &= -0.0946, & b_2^V &= 0.0446 \\
a_3^V &= 0.0342, & b_3^V &= -0.0364 \\
a_4^V &= 0.0021, & b_4^V &= 0.0081 \\
a_5^V &= -0.0019, & b_5^V &= -0.0126.
\end{aligned}
$$

Similarly, the radial velocity curve may be represented by a Fourier series. The coefficients found from a fifth order fit (minus the constant) are:

$$
\begin{aligned}
a_1^{V_r} &= 9.533, & b_1^{V_r} &= 11.306 \\
a_2^{V_r} &= -7.377, & b_2^{V_r} &= 0.535 \\
a_3^{V_r} &= 0.653, & b_3^{V_r} &= -2.991 \\
a_4^{V_r} &= 0.438, & b_4^{V_r} &= 0.834 \\
a_5^{V_r} &= -0.839, & b_5^{V_r} &= -0.164.
\end{aligned}
$$

The radial displacement is calculated by evaluating the series:

$$r = 0.181 \sum_{n=1}^{5} \frac{a_n^{V_r}}{n} \sin(2n\pi\phi) - \frac{b_n^{V_r}}{n} \cos(2n\pi\phi),$$

where the units are solar radii.

The uncertainty in the radial displacement in solar radii, from Balona (1977), is:

$$\sigma_r = \frac{pP}{2R_\odot \sqrt{N}} \sigma_{V_r}.$$

For every $K$ datum, the phase is calculated and the $V$ magnitude and radial displacement, $r$, are estimated. The angular diameter then follows from the surface-brightness calibration and a reddening law. The reddenings and total absorptions were estimated using the absorption curve parameterization of Cardelli, Clayton, and Mathis (1989). Since the extinction law is not identical along all lines of sight, it is desirable to select a parameter which can be used to scale the law appropriately. Cardelli, Clayton, and Mathis found the ratio $R_V = A_V/E_{B-V}$ worked well for this purpose. We adopt $R_V = 3.1$ and find $E_{V-K} = 2.873\ E_{B-V}$ and $A_K = 0.368\ E_{B-V}$. The form of the angular diameter relation, when expressed in units of mas, is:

$$\theta^K_{UD} = 10^{0.0504(V-K)\ +\ 0.552\ -\ 0.2K\ -\ 0.071 E_{B-V}}.$$

If $d_{kpc}$ is the distance in kiloparsecs, and $R_{AU}$ is the instantaneous stellar radius is astronomical units, then:

$$2R_{AU} = d_{kpc}\theta^K_{UD}.$$

Of course, the instantaneous stellar radius is equal to the mean stellar radius, $\bar{R}_{AU}$, plus the radial displacement, $r_{AU}$, so

$$2r_{AU} + 2\bar{R}_{AU} = d_{kpc}\theta^K_{UD}.$$

So, once we have the radial displacements and angular diameters, we can solve directly for the mean stellar radius and distance. Using a value of $E_{B-V} = 0.393$ mag from Fernie (1990), we find a distance of $0.660 \pm 0.024$ kpc and a radius of $0.2459 \pm 0.0016$ AU or $52.9 \pm 0.3$ solar radii. The distance corresponds to a true distance modulus $(m - M)_0 = 9.10 \pm 0.07$ mag. The radius is remarkably close to the estimate of 52.8 solar radii by Gieren, Barnes, and Moffett (1989) based on the period-radius relation derived using the optical Barnes-Evans technique. The precision of the estimates presented here is essentially independent of the position of the star in the instability strip. The result is consistent with, and more precise than, the distance modulus of $9.37 \pm 0.22$ mag of Gieren, Barnes, and Moffett (1993), found from application of the classical Barnes-Evans technique.

## 6. Analysis and Conclusions

How does the quality of this result compare with that from other combinations of bandpasses? The data used for the U Sgr example may be used to compare various choices of flux and color index. We select the following combinations for further examination: $V, (V - R)_0$ [Barnes-Evans]; $V, (B - V)_0$; $V, (V - K)_0$; and $K, (V - K)_0$. The radial displacement is found in the same way for each solution. For $V, (V - R)_0$ and $V, (B - V)_0$, the photometry is simultaneous and so no additional interpolation is required. For $V, (V - K)_0$ a Fourier series of fifth order was fit to the $K$ lightcurve and the following coefficients found:

$$\begin{aligned}
a_0^K &= 3.956, \\
a_1^K &= -0.009, & b_1^K &= 0.095 \\
a_2^K &= -0.022, & b_2^K &= -0.011 \\
a_3^K &= 0.017, & b_3^K &= -0.009 \\
a_4^K &= 0.004, & b_4^K &= 0.000 \\
a_5^K &= 0.004, & b_5^K &= -0.008.
\end{aligned}$$

The data used for this comparison are summarized in Table 2. The columns, from left to right, are $V$ magnitude, $(B - V)$, $(V - R)$, and $(V - K)$ color indices in magnitudes – the optical photometry being directly from from Moffett and Barnes (1984), and radial displacement in units of solar radii found from integrating the radial velocity curve already described.

Figure 2 illustrates the relative effectiveness of using the different magnitude, color index combinations to remove the effect of surface brightness change. It is clear that the introduction of the $K$ bandpass produces a dramatic reduction in the scatter of the relations and hence in the precision of the deduced distance.

Two immediate concerns for the use of this technique are: 1) Is the surface brightness color relation for Cepheids the same as for the calibrating giants, 2) are the long-period Cepheids as 'well-behaved' as their short-period counterparts, The first question is at least partially answered by Di Benedetto (1993) who finds that the $V,(V - K)_0$ relation is essentially identical for giants and supergiants for stars with the $(V - K)_0$ colors of Cepheids. It remains to be verified that Cepheids and stable supergiants have the same surface brightness color relation. The differences seen by Barnes (1980) between stable supergiants and Cepheids in the $V,(V - R)_0$ relation are worrisome, and the presence or absence of such differences in the infrared should be investigated. The second question can be addressed by studying galactic and long-period Cepheids when sufficiently precise optical and infrared photometry and radial velocities have become available.

Obvious targets for testing the precision of infrared Barnes-Evans technique are the young clusters in the Large Magellanic Cloud. The Cepheids in these clusters are predominantly short period variables, with few exceeding periods of 10 days. Their short periods allow denser phase coverage during normal observing runs. Also, the proximity of the variables to each other allows several to be captured in the same

– 13 –

infrared array field, further increasing efficiency. Some $JHK$ photometry has already been acquired for this purpose and more photometry is planned in the next observing season.

Since this technique allows the determination of distances without intermediate steps, it is natural to ask how far away it may be usefully employed. Since Welch *et al.* (1991) have demonstrated that radial velocities of precision 1 km s$^{-1}$ are possible for even short-period classical Cepheids at the distance of the LMC (50 kpc), it may be surmised that longer-period Cepheids may be studied at even greater distances. While several dwarf galaxies are found at distances intermediate between the LMC/SMC and M31, none of them has a sufficiently young population to have classical Cepheid stars. Furthermore, M31 is clearly the nearest object where many different types of distance indicators can be compared meaningfully. It is also about 5 magnitudes further away than the LMC. The most difficult observations to obtain are the high-precision radial velocities which are most frequently obtained at about 510 nm. To avoid unusual or pathological behavior and to ensure an adequate sample of objects, Cepheids with periods between 40 and 50 days are required. At the distance of M31, this translates to V = 19 mag or so. It is not, at present, possible to obtain radial velocities of the required precision at this magnitude, since 4m-class telescopes have a limit of about V = 17.5 mag for this precision level. However, by adopting slightly lower dispersion and working in a redder bandpass it is probably possible a direct distance determination to M31 or M33 with 8-10m class telescopes.

This research was supported, in part, by a Natural Sciences and Engineering Research Council (NSERC) research grant. I thank Sergio Fajardo-Acosta of SUNY at Stonybrook and Dana Backman of Franklin and Marshall College for obtaining and reducing the observation of $\alpha$ Cet. Thanks also to Pat Côté for his careful reading of an earlier draft of the paper and many useful suggestions. This research has made use of the Simbad database, operated at CDS, Strasbourg, France accessed through the Canadian Astronomy Data Centre operated by the Dominion Astrophysical Observatory.

## Figure Captions

Fig. 1 — The surface brightness-color relation at $K$ (2.2 $\mu$m) using angular diameters from Mozurkewich *et al.* (1991). This least-squares fit has a $\sigma = 0.04$ mag. The star whose angular diameter contributes most to the variance of the fit, $\delta$ And, is also the star with the smallest angular diameter of the sample. The ordinate is the $K$ magnitude of a star of a given color for an angular diameter of 1 mas. The effect of an error of in $E_{B-V}$ of 0.1 mag is shown with an arrow. Since the reddening and temperature trajectories are very similar, the derived distance to a Cepheid is very insensitive to errors in the reddening.

Fig. 2 — This plot illustrates the effectiveness which which the color can be used to remove the effect of surface brightness change on the flux. The remaining change in flux is a result of the change in cross-sectional area and hence should vary linearly with radial displacement (since magnitude is a logarithmic quantity). The precision with which the distance (or radius) can be determined is the precision of the slope of the best-fit line passing through each distribution of points. Points adjacent in phase have been connected by simple line segments. For the $V,(V-R)$ relation, the color coefficient of the Barnes-Evans technique has been used. The $V,(B-V)$ coefficient minimizes the scatter. The $V,(V-K)$ coefficient is from Di Benedetto (1993). For the $K,(V-K)$ relation, the coefficient is the derived in section 4. Photometry is from Moffett and Barnes (1984) and Laney and Stobie (1992). Radial displacement are found by integrating the radial velocities of Mermilliod, Mayor, and Burki (1987) using a projection correction factor $p = 1.36$.

Since the range in radial displacement is fixed and typically small, the precision is set by the quality of the photometry and the size of the color coefficient which, of course, will amplify or reduce the effect of errors in the colors.

Fig. 3 — The derivation of the distance and mean radius of the Cepheid from the actual distribution of data is illustrated in this plot. The radial displacement from mean radius is $r$ and $\theta_{UD}^{K}$ is the angular diameter. The percentage change in the size of the star is clearly small.

TABLE 1
Calibration Data for K Surface Brightness Relation

| Star | SpT | $\theta_{LD}$ (mas) | $\theta_{UD}^{K}$ (mas) | $K$ (mag) | $(V-K)$ (mag) | $S_K$ (mag) |
|---|---|---|---|---|---|---|
| $\alpha$ CMi | F5IV-V | 5.51 | 5.34 | $-0.671$ | 1.04 | 2.968 |
| $\beta$ Gem | K0IIIb | 8.04 | 7.75 | $-1.085$ | 2.23 | 3.361 |
| $\epsilon$ Cyg | K0III | 4.62 | 4.45 | 0.111 | 2.35 | 3.353 |
| $\delta$ And | K3III | 4.12 | 3.96 | 0.430 | 2.85 | 3.418 |
| $\alpha$ Cas | K0IIIa | 5.64 | 5.43 | $-0.250$ | 2.48 | 3.424 |
| $\alpha$ Ari | K2IIIab | 6.85 | 6.59 | $-0.636$ | 2.64 | 3.458 |
| $\gamma^1$ And | K3IIb | 7.84 | 7.52 | $-0.810$ | 3.07 | 3.572 |
| $\alpha$ Tau | K5III | 21.21 | 20.29 | $-2.850$ | 3.71 | 3.686 |
| $\beta$ And | M0IIIa | 13.81 | 13.19 | $-1.860$ | 3.91 | 3.742 |
| $\alpha$ Cet | M1.5IIIa | 13.23 | 12.62 | $-1.697$ | 4.23 | 3.808 |
| $\beta$ Peg | M2.5II-III | 17.98 | 17.11 | $-2.246$ | 4.67 | 3.920 |

TABLE 2
U Sgr Photometry and Radial Displacements

| $V$ (mag) | $(B-V)$ (mag) | $(V-R)$ (mag) | $(V-K)$ (mag) | $r$ ($R_\odot$) |
|---|---|---|---|---|
| 6.389 | 0.947 | 0.834 | 2.408 | −3.03 |
| 6.374 | 0.929 | 0.840 | 2.398 | −2.97 |
| 6.338 | 0.913 | 0.802 | 2.395 | −2.39 |
| 6.353 | 0.910 | 0.812 | 2.419 | −2.15 |
| 6.400 | 0.943 | 0.804 | 2.483 | −1.00 |
| 6.390 | 0.945 | 0.808 | 2.473 | −0.92 |
| 6.420 | 0.945 | 0.823 | 2.503 | −1.02 |
| 6.500 | 1.010 | 0.913 | 2.605 | 0.28 |
| 6.496 | 1.024 | 0.866 | 2.604 | 0.41 |
| 6.589 | 1.078 | 0.927 | 2.704 | 1.22 |
| 6.601 | 1.095 | 0.950 | 2.714 | 1.43 |
| 6.609 | 1.095 | 0.952 | 2.722 | 1.43 |
| 6.617 | 1.118 | 0.942 | 2.732 | 1.72 |
| 6.610 | 1.123 | 0.944 | 2.729 | 1.84 |
| 6.649 | 1.153 | 0.948 | 2.779 | 2.01 |
| 6.670 | 1.149 | 0.927 | 2.804 | 2.08 |
| 6.727 | 1.187 | 0.949 | 2.854 | 2.14 |
| 6.720 | 1.181 | 0.953 | 2.843 | 2.14 |
| 6.777 | 1.198 | 0.976 | 2.871 | 2.11 |
| 6.812 | 1.200 | 1.005 | 2.895 | 2.09 |
| 6.876 | 1.228 | 1.016 | 2.939 | 2.04 |
| 6.881 | 1.232 | 0.990 | 2.938 | 2.02 |
| 6.924 | 1.252 | 1.018 | 2.975 | 1.88 |
| 6.921 | 1.259 | 1.046 | 2.972 | 1.85 |
| 6.983 | 1.290 | 1.003 | 3.027 | 1.46 |
| 7.014 | 1.286 | 1.022 | 3.035 | 1.05 |
| 7.045 | 1.320 | 1.042 | 3.010 | 0.00 |
| 7.060 | 1.294 | 1.059 | 3.022 | −0.09 |
| 7.043 | 1.275 | 1.041 | 2.990 | −0.51 |
| 7.045 | 1.269 | 1.041 | 2.982 | −0.91 |
| 6.987 | 1.210 | 0.987 | 2.902 | −2.29 |
| 6.959 | 1.180 | 0.972 | 2.871 | −2.65 |
| 6.811 | 1.139 | 0.943 | 2.725 | −3.20 |
| 6.643 | 1.062 | 0.908 | 2.590 | −3.56 |
| 6.551 | 1.007 | 0.855 | 2.530 | −3.41 |
| 6.552 | 0.991 | 0.856 | 2.528 | −3.44 |

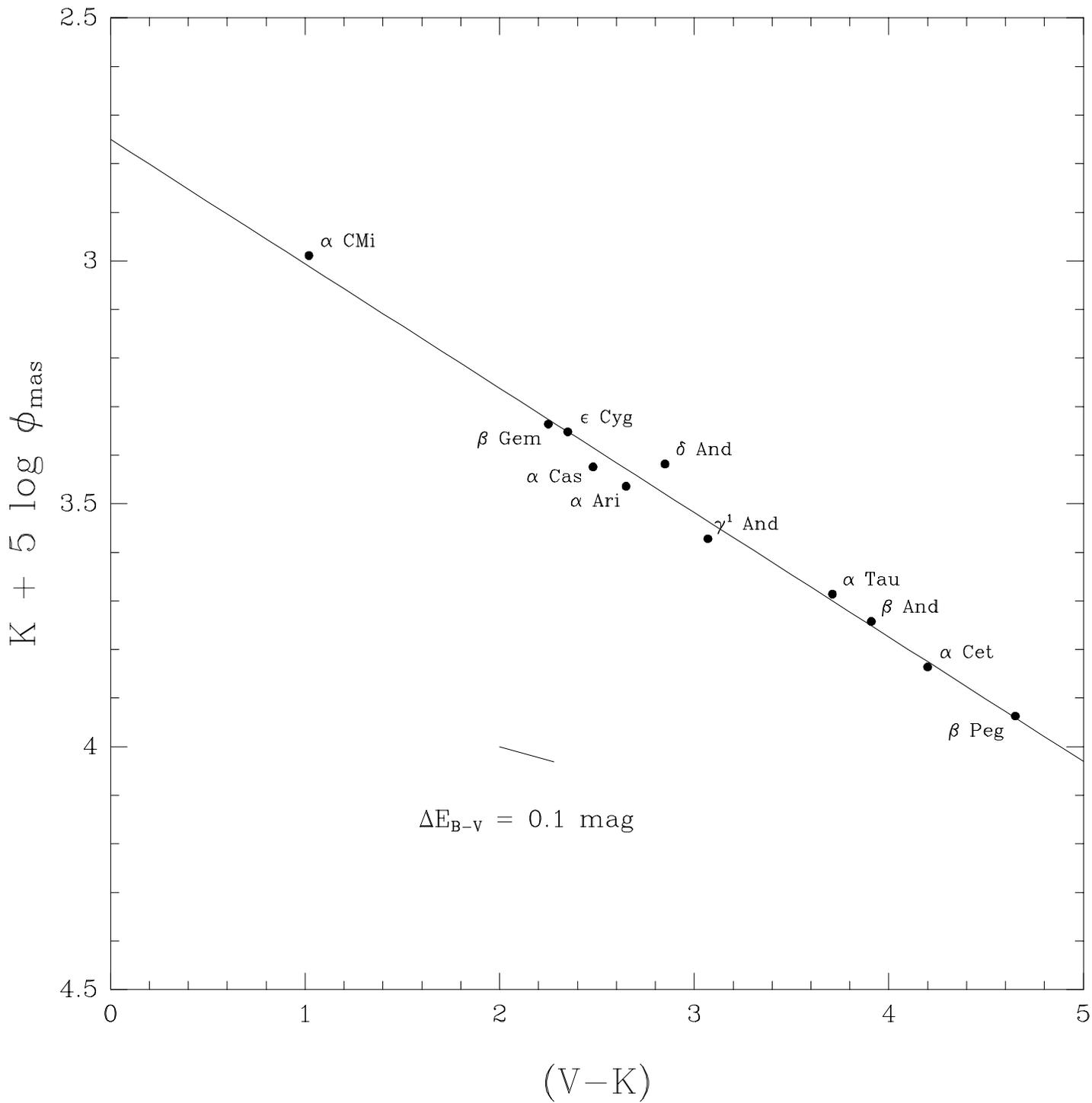

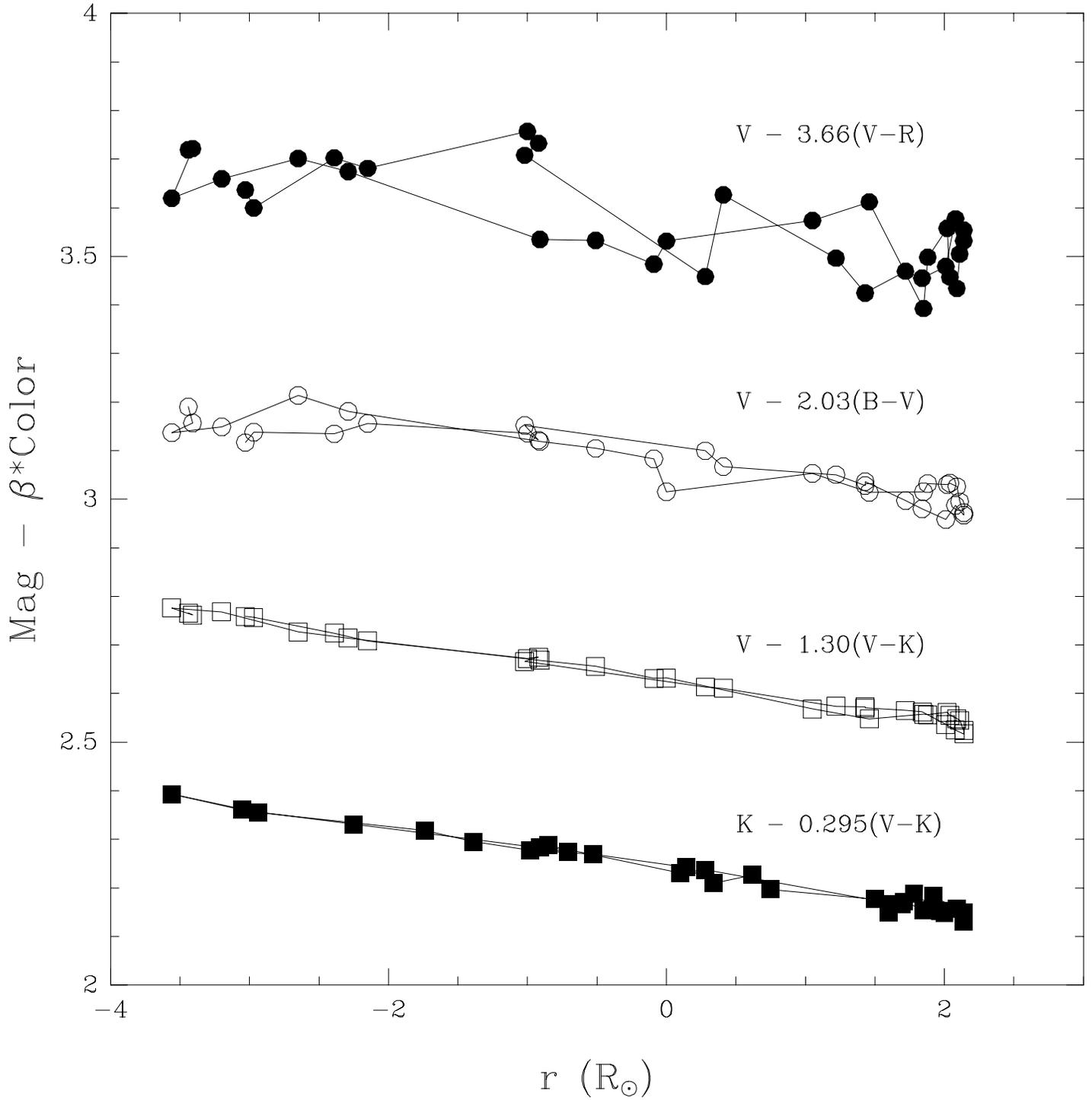

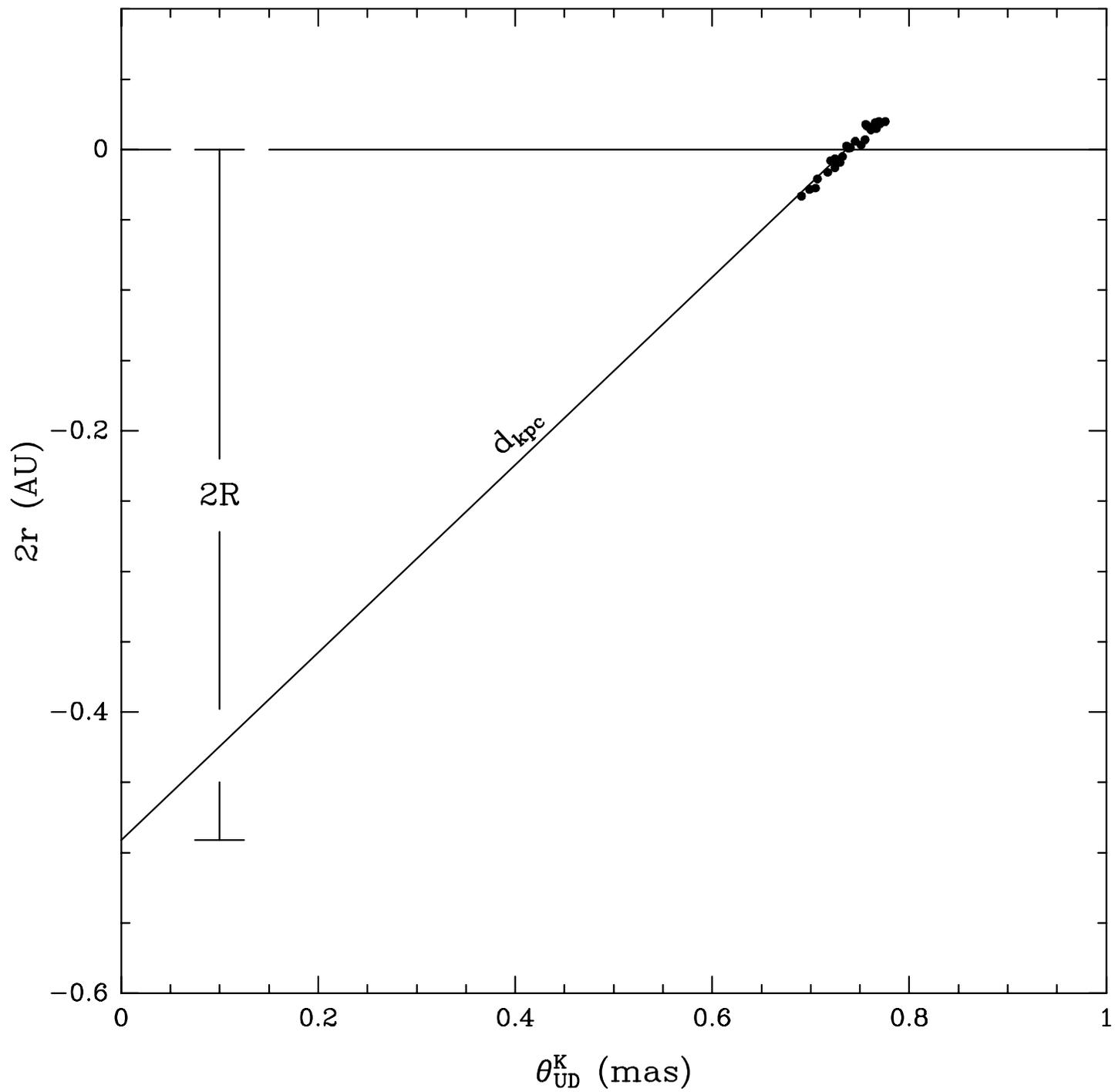